\documentclass{article}
\usepackage{spconf,amsmath,graphicx,multirow,booktabs}

\title{GMM-ResNext: Combining Generative and Discriminative Models for Speaker Verification}
\name{Hui Yan, Zhenchun Lei*\thanks{*Corresponding author}, Changhong Liu, Yong Zhou}
\address{School of Computer and Information Engineering, Jiangxi Normal University, Nanchang, China}
\begin{document}
\ninept
\maketitle
\begin{abstract}
With the development of deep learning, many different network architectures have been explored in speaker verification. However, most network architectures rely on a single deep learning architecture, and hybrid networks combining different architectures have been little studied in ASV tasks. In this paper, we propose the GMM-ResNext model for speaker verification. Conventional GMM does not consider the score distribution of each frame feature over all Gaussian components and ignores the relationship between neighboring speech frames. So, we extract the log Gaussian probability features based on the raw acoustic features and use ResNext-based network as the backbone to extract the speaker embedding. GMM-ResNext combines Generative and Discriminative Models to improve the generalization ability of deep learning models and allows one to more easily specify meaningful priors on model parameters. A two-path GMM-ResNext model based on two gender-related GMMs has also been proposed. The Experimental results show that the proposed GMM-ResNext achieves relative improvements of 48.1\% and 11.3\% in EER compared with ResNet34 and ECAPA-TDNN on VoxCeleb1-O test set.
\end{abstract}
\section{Introduction}
\label{sec:intro}

The task of Automatic Speaker Verification (ASV) \cite{2021Speaker} is to verify the identity of speakers by using speaker speech as feature. For two given utterances, a typical ASV system can extract speaker embeddings of them and automatically determine whether they belong to the same speaker or not. The procedure of a modern ASV system generally consists of acoustic feature extraction, speaker embedding extraction and similarity scoring. The purpose of acoustic feature extraction is to transform a waveform of speech into acoustic features, such as Filter-Banks, Mel Frequency Cepstral Coefficients (MFCCs) and spectrograms. Speaker embedding extraction is to extract fixed-length speaker embeddings from variable-length utterances. Similarity scoring aims to calculate the similarity between test speaker embedding and enrollment speaker embedding.

The conventional generative models such as Gaussian mixture model-Universal Background Model (GMM-UBM) \cite{2000Speaker} and i-vector \cite{Dehak2011Front} with Probabilistic Linear Discriminant Analysis (PLDA) \cite{10.1007/11744085_41} were main method in the field of ASV. With the development of deep learning techniques, these models have been gradually replaced by Deep Neural networks (DNNS). Now, more and more DNN-based models were applied to ASV tasks. Especially r-vector structures based on convolution neural network (CNN) and x-vector architectures based on Time Delay Neural Network (TDNN) shown remarkable performance. For CNN-based ASV systems, Zeinali et al. \cite{2019BUT} firstly used Residual Networks (ResNet) \cite{7780459} in image recognition as speaker embedding extractor in VoxSRC 2019. Liu et al. \cite{liu22f_interspeech} proposed ResNet based on sequential and parallel feature attention fusion mechanism, which uses attention mechanism to learn fusion weights based on feature content, thereby dynamically integrating identity mapping features and residual learning features. Chen et al. \cite{chen23o_interspeech} proposed an enhanced Res2Net, which uses  attention fusion module to fuse features in a residual block to  extract local signals, and fuse features of different scales to  aggregate global signals. For TDNN-based ASV systems, x-vector proposed by Snyder et al. \cite{8461375} employed TDNN to map variable-length speech to fixed-length speaker embedding for the first time, and used PLDA back-end model to compare similarity of a pair of speaker embeddings. Desplanques et al. \cite{desplanques20_interspeech} proposed ECAPA-TDNN based on Squeeze-Excitation (SE) \cite{8578843} module and Res2Net\cite{2019Res2Net}, which achieves the equal error rates of less than 1$\%$ in VoxCeleb-O test set. Thienpondt et al. \cite{2021Integrating} proposed ECAPA CNN-TDNN, which introduces a 2-D convolution into ECAPA-TDNN to transfer some strong characteristics of ResNet to this hybrid CNN-TDNN architecture, and achieves better results on VoxCeleb-O test set. Zhao et al. \cite{10095051} proposed to segment the input spectral map into several frequency bands and use progressive channel fusion strategy to gradually fuse these bands to improve the ECAPA-TDNN.

Although the  embedding extractors mentioned above show excellent performance in ASV task, they only rely on a single deep learning architecture. In the filed of ASV, Alam et al. \cite{10096040} proposed  a hybrid neural network(HNN) architecture with Cross- and Self-module Attention pooling mechanisms for speaker verification.  Wang et al. \cite{9687918} proposed MACCIF-TDNN, which is a series of residual networks and transformers. Wang et al. \cite{wang23i_interspeech} proposed a Parallel-coupled TDNN/Transformer Network (p-vectors) to  replace the serial hybrid networks. In the field of speech deepfake detection, Lei et al. \cite{2021Lei} proposed GMM-Transformer, which is a fusion of Gaussian Mixture Model (GMM) and deep neural network, and achieves excellent detection effect in the speech deepfake detection logical access task. Wen et al. \cite{wen22_interspeech} also proposed multi-branch GMM-MobileNet model based on different data augmentation and attack methods. However, hybrid networks that combine conventional machine learning methods with popular deep learning methods have been rarely studied in ASV tasks. 

\begin{figure*}[t]

\begin{minipage}[b]{1.0\linewidth}
  \centering
  \centerline{\includegraphics[width=15.0cm]{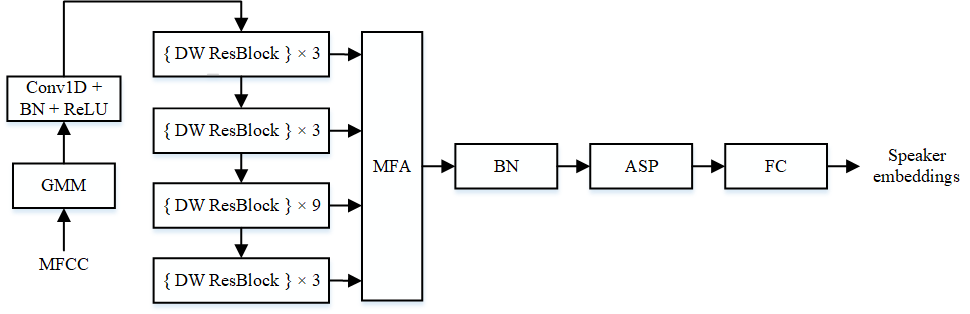}}
\end{minipage}

\caption{The overall architecture of GMM-ResNext}
\label{fig:overture}
\end{figure*}
The Gaussian Mixture Model accumulates the scores on all frames independently, and does not separately consider the scores of feature frames on each Gaussian component. In addition, the GMM ignores the relationship between adjacent speech frames along the time dimension. In this paper, we propose the GMM-ResNext that applies the Gaussian probability features as input for speaker verification. Then our proposed GMM-ResNext concatenates the output feature maps from last layer in each stage to aggregate the multi-layer representations before final pooling. On the other hand, the specificity of male speech features and female speeches features is useful for modeling the feature distribution of speech. Therefore we propose the dual-path GMM-ResNext(dGMM-ResNext) based on the different genders.
 
\section{ Log Gaussian Probability Feature}
\label{sec:LGP}

The GMM is a conventional speaker recognition classifier. For a speech feature vector, the  GMM sums the probability density values of N Gaussian components, but does not consider the score distribution of each Gaussian component separately. The relationship between adjacent speech frames is ignored. For the information distributions of speeches of different speakers are different in feature space, their score distributions on all Gaussian components are also different. Therefore, this score distribution information is helpful for speaker verification. The GMM takes the raw acoustic feature as input and outputs the Log Gaussian Probability(LGP) feature. For a D dimensional input feature x (MFCC in our experiments), the element $y_{i}$ of the LGP feature $y$ is defined as:
\begin{align}
	y_{i} &= \log p_{i}(x)  \nonumber\\ 
              &= \log \frac{1}{(2\pi)^{D/2}|\Sigma_{i}|^{1/2}}\exp\{-\frac{1}{2}(x-\mu_{i})'\Sigma_{i}^{-1}(x-\mu_{i})\}\nonumber\\
              &= -\frac{1}{2}x'\Sigma_{i}^{-1}x + x'\Sigma_{i}^{-1}\mu_{i}+C
\end{align}
where $p_{i}(x)$ is the probability density function of the i-th Gaussian component,which is parameterized by a $D \times 1$ mean vector, $\mu_{i}$ and a $D \times D$ covariance matrix, $\Sigma_{i}$.

In order to reduce computation, the constant term $C$ can be removed. After that, the mean $mean_{y_{i}}$ and standard deviation $std_{y_{i}}$ of features in training data are computed, which are used for mean and variance normalization.
\begin{align}
	y_{i}' &= \frac{y_{i}-mean_{y_{i}}}{std_{y_{i}}}
\end{align}

\section{proposed method}
\label{sec:gmm-resnet}

In this section, we describe details of the proposed GMM-ResNext model and dual-Path GMM-ResNext.

\subsection{GMM-ResNext}
\label{ssec:gmmnet}

The CNN-based model is widely used as the backbone network of various machine learning tasks because of its stronger feature extraction ability. However, with the increase of the number of layers in the network, the problem of gradient explosion or gradient disappearance or even network degradation will appear in the training process. In order to solve the above problems, the ResNet \cite{7780459} employs shortcut connections to fusion the identity mapping features and the residual learning features to improve the stability of the model. In ResNet, the number of parameters in the model is mainly adjusted by depth and width. In practice, we usually find that directly increasing the depth or width of the model is ineffective, and easily cause overfitting problems due to the large number of parameters. The ResNext \cite{2017Aggregated} that includes the grouped convolutional layer called a split-transform-merge strategy in blocks is designed to easy this problem and improve the representation ability of ResNet model. The ResNext has achieved reliable performance in SV \cite{2020ResNeXt}, so we proposed the GMM-ResNext model that fuses the GMM and ResNext for speaker verification. The overall architecture is shown in Figure~\ref{fig:overture}. 

The proposed GMM-ResNext model consists of a GMM module, a ResNext backbone with four residual stages, An attention statistics pooling (ASP) layer with Multi-layer Feature Aggregation (MFA), and a fully connected layer. Encouraged by \cite{2022A} proposed design principles, we adjusted the number of blocks in each stage from (3, 4, 6, 3) to (3, 3, 9, 3), which is different from the original stage ratio in the residual network. BN stands for Batch Normalization and the Rectified Linear Units (ReLU) is the non-linear activation function.

\begin{figure}[t]
\begin{minipage}[b]{1.0\linewidth}
  \centering
  \centerline{\includegraphics[width=3.5cm]{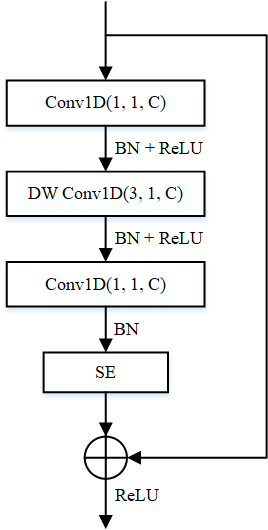}}
\end{minipage}

\caption{The architecture of DW ResBlock. The numbers in parentheses refer to kernel size, stride, and number of channel.}
\label{fig:dwres}
\end{figure}

To further reduce the number of parameters in the model, we use depthwise convolution instead of grouped convolution in ResNext, which is a special case of grouped convolution where the number of groups equals the number of channels. The structure of the DW-ResBlock module is shown in Figure~\ref{fig:dwres}. The module consists of two convolutional layers with filter size of 1, 1D depthwise convolutional layer with filter size of 3, SE block, and BN and ReLu after the convolutional layer. In the SE Block, the dimension of the bottleneck is set to 1/4 of the number of input channel.

Previous studies \cite{gao19_interspeech} have shown that the shallow feature maps in deep neural networks also facilitate the extraction of more robust speaker embeddings. In \cite{desplanques20_interspeech}, the ECAPA-TDNN concatenates the output features of all SE-Res2Net blocks, and then uses a dense layer to process the connected feature information to generate the input of the pool layer. Zhang et al. \cite{zhang22h_interspeech} proposed MFA-Conformer model that concatenates the output features of each Conformer block, and then feed them into a LayerNorm layer. This aggregation method results in a obvious performance improvement. Influenced by these studies, we also concatenates the output feature maps of all stages to generate multi-level input features of the pooled layer before a BN layer.
\begin{align}
	H &= BatchNorm(Concat(h_{1}, h_{2}, ..., h_{L}))
\end{align}
Where $h_{i}\in R^{C\times T}$ represents the output feature of last layer in each stage and $H\in R^{C\times T}$. $Concat$ is a concatenation operation. $L$ is the number of stages and $D = C\times{T}$.

After aggregating the output features of the different residual blocks, we use the ASP to capture the weight coefficients that represents the importance of each frame. The output of the pooling layer is given by concatenating the weighted mean vector and the weighted standard deviation calculated by the weight coefficient. Finally, We use the full connection layer to project the pooled vector into a low-dimensional speaker embedding. 

\subsection{Dual-path GMM-ResNext}
\label{ssec:tgmmnet}

Improving the generalization ability of speaker verification is a
great challenge. The speech from speakers of different genders has different feature distribution. The  conventional GMM describes the common information distribution. But the GMM does not pay attention to the difference between male and female speeches, which is useful to model information distribution of speech from different genders. So we propose a dual-path architecture with ResNext (dGMM-ResNext) for speaker verification, which is constructed according to two GMMs trained by male speeches and female speeches.

\begin{figure}[t]

\begin{minipage}[b]{1.0\linewidth}
  \centering
  \centerline{\includegraphics[width=8.5cm]{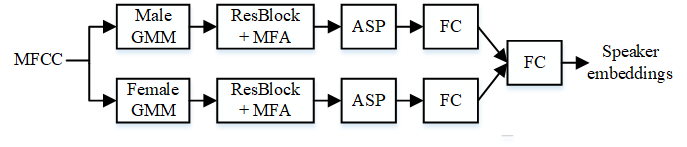}}
\end{minipage}

\caption{The architecture of dual-path GMM-ResNext based on male GMM and female GMM.}
\label{fig:DP}
\end{figure}

Figure~\ref{fig:DP} shows the dual-path GMM-ResNext which contains two networks with same architecture. The LGP features are extracted on two GMMs, which are trained on male speeches and female speeches from training set. The embedding vectors from two paths are concatenated and inputted to the fully connected layer. Finally, the speaker embedding are output from the last fully connected layer.

Multi-step training scheme \cite{jung18b_interspeech} is usually used to solve the problem of model overfitting. Considering the large number of model parameters, a two-step training method is adopted to improve the robustness of model. In the first step, the two branches of the dual-path ResNext model are trained independently using the AAM-softmax loss function. The training method is the same as that of the one-path model. In the second step, we remove their classifiers and concatenate the speaker embeddings of the two branches into one fully connected layer to generate the final speaker embedding. Then we freeze the parameters of all layers of the dual-dual ResNext except the classifier to train the fully connected layer and the classifier.

\section{Experimental setup}
\label{sec:setup}

\subsection{Dataset}
\label{ssec:da}

The experiments are conducted on VoxCeleb1 \cite{2017VoxCeleb} and VoxCeleb2 \cite{2018VoxCeleb2} dataset. They include a development set and a test set. VoxCeleb2 development set is used for training, which consists of 109,200,9 utterances from 599,4 speakers. The whole VoxCeleb1 dataset is used as the testing data, which contains over 100,000 utterances from 125,1 speakers. The performance of all models are evaluated on three different test trials, namely VoxCeleb1-O, VoxCeleb1-E and VoxCeleb1-H. In addition, to increase the diversity of the original training data, we employed data augmentation during training, adding noise using the MUSAN dataset \cite{2015MUSAN}, simulating reverberation using the RIR dataset \cite{2017A}.

\subsection{Implementation details}
\label{ssec:de}

For fair comparison, we re-implemented r-vector in \cite{2019BUT} and ECAPA-TDNN in \cite{desplanques20_interspeech} as the baselines. The r-vector includes ResNet18 and ResNet34 with channels of residual blocks set as \{32,64,128,256\}. ECAPA-TDNN consists of a basic model with 512 channels and a large model with 1024 channels.
\begin{table*}[t]

  \caption{EER and MinDCF results of all models on the standard VoxCeleb1 test set.}
  \label{tab:EERDCF}
  \centering
  \begin{tabular}{ccccccccc}
    \toprule
    \multicolumn{3}{c}{\multirow{2}{*}{Model}} 
    &\multicolumn{2}{c}{VoxCeleb1-O} 
    &\multicolumn{2}{c}{VoxCeleb1-E} 
    &\multicolumn{2}{c}{VoxCeleb1-H}\\
    
    \cmidrule(lr){4-5} \cmidrule(lr){6-7} 	\cmidrule(lr){8-9}
    
    \multicolumn{3}{c}{}
    &\multicolumn{1}{c}{EER(\%)}&\multicolumn{1}{c}{MinDCF}
    &\multicolumn{1}{c}{EER(\%)}&\multicolumn{1}{c}{MinDCF}
    &\multicolumn{1}{c}{EER(\%)}&\multicolumn{1}{c}{MinDCF} \\
			
    \midrule

    \multicolumn{3}{c}{ResNet18\cite{2019BUT}}
    &1.97&0.2502&2.15&0.2445&3.94&0.3659 \\
    \multicolumn{3}{c}{ResNet34\cite{2019BUT}}
    &1.81&0.2108&1.92&0.2183&3.46&0.3379 \\
    \multicolumn{3}{c}{ECAPA-TDNN(512)\cite{desplanques20_interspeech}}
    &1.22&0.1455&1.41&0.1609&2.59&0.2555 \\
    \multicolumn{3}{c}{ECAPA-TDNN(1024)\cite{desplanques20_interspeech}}
    &1.06&0.1310&1.32&0.1495&2.49&0.2532 \\
    \midrule
    
    \multicolumn{3}{c}{CA-HNN\cite{10096040}}&1.38&-&1.62&-&2.86&- \\
    \multicolumn{3}{c}{CSA-HNN\cite{10096040}}&1.32&-&1.53&-&2.79&- \\
    \multicolumn{3}{c}{MACCIF-TDNN\cite{9687918}}&1.19&0.148&1.47&0.158&2.48&0.235 \\
    \multicolumn{3}{c}{P-vector\cite{wang23i_interspeech}}&0.85&0.1199&1.11&0.1201&2.11&0.2081 \\
    
    \midrule
    \multicolumn{3}{c}{GMM-ResNext(256)}&1.35&0.1430&1.54&0.1741&2.84&0.2697 \\
    \multicolumn{3}{c}{GMM-ResNext(512)}&\bf0.96&\bf0.1168&\bf1.20&\bf0.1424&\bf2.31&\bf0.2247 \\
    \multicolumn{3}{c}{dGMM-ResNext(256)}&1.18&0.1478&1.33&0.1537&2.48&0.2387 \\
    \multicolumn{3}{c}{dGMM-ResNext(512)}&\bf0.94&\bf0.1100&\bf1.13&\bf0.1298&\bf2.18&\bf0.2207 \\
    
    \bottomrule
  \end{tabular}
  
\end{table*}
The PyTorch framework is used to implement the baseline and the proposed models. A fixed length 2-second segments are extracted randomly from each utterance. We use 80-dimensional MFCCs with a window length of 25 ms and a frame shift of 10 ms. Mean normalization is applied to the MFCCs features before input network. The GMM implemented by MSR Identity Toolbox\cite{2013MSR} was trained for 30 iterations. Adam optimizer with an initial learning rate of
0.001 is used during the training process.  The learning rate is reduced by 3$\%$ every one epoch. We use additive margin softmax (AAM-softmax) \cite{Deng_2019_CVPR} loss with a margin of 0.2 and a scale factor of 30 to train all models. In order to avoid overfitting, the weight decay is set to 2e-5. In the ASP, the dimension of the bottleneck is set to 128. The batch size is 200 and the size of speaker embeddings is 256. All models are trained for 100 epochs.

In the evaluation phase, we use the cosine similarity between embeddings for scoring. the Equal Error Rate (EER) and minimum Detection Cost Function (minDCF) with $p_{target} = 0.01$ and $C_{FA} = C_{Miss} = 0.01$ will be reported for performance evalution.

\section{RESULTS AND ANALYSIS}
\label{sec:results}
\subsection{Results on VoxCeleb test set}
Table~\ref{tab:EERDCF} shows the EER and the MinDCf of the baselines and the proposed models on the VoxCeleb1-O, VoxCeleb1-E and VoxCeleb1-H datasets. Compared with GMM-ResNext(256), the GMM-ResNext(512) obtains the relative improvements in EER by 28.8$\%$, 22.1$\%$, 18.7$\%$ and in MinDCF by 18.3$\%$, 18.2$\%$, 16.7$\%$. The dGMM-ResNext(512) also obtains similar performance improvements compared with dGMM-ResNext(256). It can be observed that when the number of Gaussian components increase, the GMM-ResNext can achieve better performance. This may be due to the fact that Gaussian probability features contain more distinguishing speaker information.

Compared with GMM-ResNext(256), the dGMM-ResNext(256) obtains the relative improvements in EER by 12.6$\%$, 13.6$\%$, 12.7$\%$ on the VoxCeleb1 test set. The dGMM-ResNext(512) further improves improvement compared with GMM-ResNext(512). The dGMM-ResNext(512) obtains the best results, which relatively reduces the EER by 48.1$\%$, 41.1$\%$, 37.0$\%$ and the MinDCF by 47.8$\%$, 40.5$\%$, 34.7$\%$  compared with the ResNet34, and relatively reduces the EER by 11.3$\%$, 14.4$\%$, 12.4$\%$ and the MinDCF by 16.0$\%$, 13.2$\%$, 12.8$\%$ compared with the ECAPA-TDNN(1024). 

Compared with other HNN systems, such as CA-HNN, CSA-HNN and MACCIF-TDNN, the dGMM-ResNext(512) also achieves competitive results. But the performance of the dGMM-ResNext(512) is worse than P-vector. The reason may be that P-vector leverages the convolutional operations
and the attention mechanism to interact and aggregate the local and global information.
 
\subsection{Ablation experiments}
To evaluate the role of key modules, we conduct ablation experiments to study the effect of each modules contributing to performance improvements. Table~\ref{tab:albation} shows the results of the ablation experiments on VoxCeleb1-O test set. The first line is the results of the GMM-ResNext(512). In the second line, we remove the GMM layer and use the MFCC feature instead of the log-Gaussian probabilistic feature as the input of the network. In the third line, the MFA layer is removed and only the output feature map of the last residual block is used. The results in the fourth line are the dGMM-ResNext(512). In the last line, we does not use the two step training strategy.  It can be observed from the experiment results that both GMM layer and multi-layer feature aggregation module play a key role in improving the performance. The GMM layer reduces EER and minDCF by 21.3$\%$ and 19.3$\%$ relatively. And aggregating the output of all blocks brings 16.5$\%$ and 21.2$\%$ relative improvement in EER and minDCF. At the same time, the two step training strategy further improves the generalization ability of the dGMM-ResNext.
\begin{table}[h]

  \caption{Ablation experiments on VoxCeleb1-O. 2S: The two step training strategy. }
  \label{tab:albation}
  \centering
  \begin{tabular}{cccc}
    \toprule
    \multicolumn{2}{c}{\multirow{2}{*}{Model}} 
    &\multicolumn{2}{c}{Vox1Celeb-O} \\
    
    \cmidrule(lr){3-4}
    
    \multicolumn{2}{c}{}
    &\multicolumn{1}{c}{EER(\%)}&\multicolumn{1}{c}{MinDCF} \\
			
    \midrule
    
    \multicolumn{2}{c}{ResNext(512)}&0.96&0.1168 \\
    \multicolumn{2}{c}{w/o GMM)}&1.22&0.1447 \\
    \multicolumn{2}{c}{w/o MFA)}&1.15&0.1483 \\
    \midrule
    \multicolumn{2}{c}{dGMM-ResNext(512)}&0.94&0.1100 \\
    \multicolumn{2}{c}{w/o 2S}&1.04&0.1172 \\
    
    \bottomrule
  \end{tabular}
  
\end{table}

\section{ CONCLUSION}
\label{sec:concl}

In this paper, we proposed the GMM-ResNext model that combines conventional machine learning methods with deep learning methods for speaker verification. First, the proposed model uses Gaussian probability features as the input of the residual network. Then, the output feature maps of the four stages are aggregated to integrate the multi-level features, so as to obtain more distinguishable speaker information. We also proposed the dual-path GMM-ResNext model based on different genders to improve the generalization ability of the model. Experiments on the VoxCeleb dataset show that the proposed dGMM-ResNext model is significantly superior to the currently popular ResNet and ECAPA-TDNN models. In future work, we will explore new network architectures fused with Gaussian mixture model.

\section{Acknowledgements}

This work is supported by National Natural Science Foundation of China (62067004).
\bibliographystyle{IEEEbib}
\bibliography{strings,refs} 

\end{document}